\documentclass[twocolumn]{elsart3}

\usepackage{graphicx}
\usepackage{subfigure}

\begin{document}

\begin{frontmatter}



\title{Test Beam Results of Geometry Optimized Hybrid Pixel Detectors}

\author[adr:wtal]{K.-H.Becks}
\author[adr:wtal]{P.Gerlach}
\ead{peter.gerlach@physik.uni-wuppertal.de}
\author[adr:wtal]{C.Grah\thanksref{presCG}}
\author[adr:wtal]{P.M{\"a}ttig}
\author[mpi]{T.\,Rohe\thanksref{presTR}}

\address[adr:wtal]{Bergische Unversit{\"a}t Wuppertal. Germany}
\address[mpi]{Max--Planck-Institut f{\"u}r Physik, Munich, Germany}
\thanks[presCG]{Present assress: DESY-Zeuthen, Berlin, Germany}
\thanks[presTR]{Present address: Paul Scherrer Institut, Villigen,
Switzerland}

\begin{abstract}
The Multi-Chip-Module-Deposited (MCM-D) technique has been used to build hybrid pixel detector assemblies.
This paper summarises the results of an analysis of data obtained in a test beam campaign at CERN.
Here, single chip hybrids made of ATLAS pixel prototype read--out electronics and special sensor tiles were used. They were prepared by the Fraunhofer Institut f\"ur Zuverl\"assigkeit und Mikrointegration, IZM, Berlin, Germany.
The sensors feature an optimized sensor geometry called ``equal sized bricked''. 
This design enhances the spatial resolution for double hits in the long direction of the sensor cells.
\end{abstract}

\end{frontmatter}

\section{Introduction}
\label{sec:intro}
The Multi-Chip-Module-Deposited (MCM-D) technique was studied as part of the R\&D work for the 
ATLAS pixel detector project (for the latter, see several contributions in this proceedings).
Although for this system a more conservative approach has been chosen using a flexible 
circuit glued on the back side of the sensor and wire--bonds for the i/o--connections of the electronics,
the MCM-D technique allows for some unique features. As bump--bonding is used for all the connections,
the final modules are robust and easy to handle. 

The realization of the connection structures (see section \ref{sec:mcmd}) enhances the bump-bonding process, 
which is needed for hybrid pixel detectors.
This reduces the amount of manual assembly steps and moves the module assembly 
into the semiconductor industry environment using wafer treatment technologies.
The hybrids analysed here where fabricated by the Fraunhofer Institut f{\"u}r Zuverl{\"a}ssigkeit und Mikrointegration, 
IZM (Berlin, Germany).
The manufacturing at IZM included the build up of the thin--film layers, 
the bumping with Pb/Sn, and the flip-chip assembly. A detailed description of the same bumping process, 
used for the ATLAS pixel modules, is presented by IZM in these proceedings.

\section{MCM--D technique}
\label{sec:mcmd}
By using thin film technologies like sputtering, electro--plating,
and spin-on techniques the interconnect system can be built in an
approximately 30\,$\mu$m thin multi--layer system directly on the sensor
substrate at wafer level. At first a spin--on technique is used to
deposit a thin layer of photosensitive polymer onto a prepared
silicon sensor. The polymer is benzocyclobutene in a photosensitive
formulation, called Photo--BCB and distributed by Chemical, which
allows for the use of standard techniques of the semiconductor industry
to structure the layer. Via openings of diameters down to
22\,$\mu$m have been produced with low failure rate
\cite{mcmdpixel}.

The subsequent Cu layer is deposited by electro--plating to achieve
a sufficient layer thickness for power distribution within the
module. By using a combination of sputtering, photo resist
patterning, and electro--plating, minimal structure sizes down to
15\,$\mu$m line width and 15\,$\mu$m gap have been realized for
the first time in our current design generation.

\begin{figure}[b]
	\centering
		\includegraphics[clip=true,width=\columnwidth]{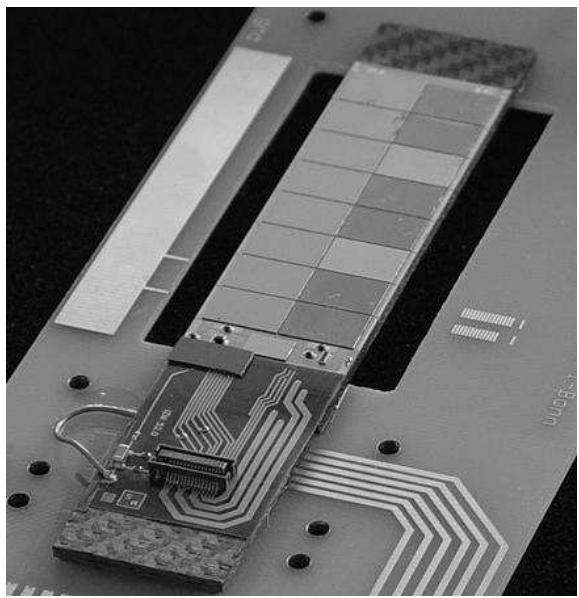}
	\caption{Picture of a full-featured MCM-D module.}
	\label{fig:mod}
\end{figure}
Figure \ref{fig:mod} shows a picture of a module
built in MCM--D. Four Cu layers of $\sim 3.5\mu\mathrm{m}$ thickness
are sufficient to fulfil the ATLAS requirements for voltage drops
on the two supply voltages to be less than 100\,mV. 
The length of the supply lines extend over the full
length of a module with a minimal width of $\approx 320\,\mu$m. 
The signal bus system is designed in a microstrip configuration 
with outstanding performance and low crosstalk \cite{becks}. 
The performance of such a module has been described in \cite{NSS04}.

\section{Equal--Sized--Bricked Sensors}
\begin{figure}[t]
	\centering
		\includegraphics[clip=true,width=\columnwidth]{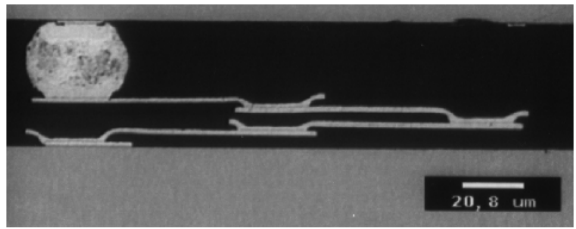}
	\caption{Cross section of a feed-through (IZM).}
	\label{fig:feedthrough}
\end{figure}

Building up the module's interconnection structures on the sensor means 
that every sensor cell has to be connected to it's corresponding readout 
electronics through the thin-film layers. A feed-through structure 
establishing this connection is shown in cross section in figure \ref{fig:feedthrough}.
The performance of the system is not affected by the additional capacitance and resistance 
of the feed-throughs, as shown in \cite{becks,mcmdpixel}.\\
The feed-thoughs decouple the sensor geometry from the layout of the electronics.

In conventional hybrid detectors the sensor geometry has to follow the size of the 
electronic cells. In addition, gaps at chip borders need to be filled by enlarged or
coupled sensor cells. The feed-throughs allow for an optimisation of the sensor layout 
and a matching to the electronic cells. For the studies done here, 
an ``equal--sized--bricked'' design has been implemented making use of
two main ideas.

The first is to divide the total area covered by the electronic ASICs 
(including dead rims and gaps) by the number of readout channels. 
The resulting dimensions are taken for the sensor cell.
The readout electronics taken from an ATLAS pixel preproduction has a cell size of 
$400 \times 50 \mu\mathrm{m}^2$. 
Each chip is providing 2880 readout channels, resulting in a total area of 
$16.4 \times 60.4 \mathrm{mm}^2$ to be covered by the sensor. 
This includes additional area to cover the $100 \mu\mathrm{m}$ wide dead rim 
of the electronics and a gap of $200 \mu\mathrm{m}$ between chips.
The resulting size of a sensor cell for an equal sized design is 
$422.22 \times 52.25 \mu\mathrm{m}^2$.

The second is to ``shift'' even rows of sensor cells by a quarter of a cell 
length to one side and odd rows by the same amount to the right. 
The result looks like a bricked wall and features an equal sized segmentation, 
which eases the spatial reconstruction and an enhanced spatial resolution for double
hits composed by cells in adjacent rows.

In order to keep the ability to test the n-on-n sensors before the flip--chip
hybridisation, two different geometries of the bias--grid were implemented.

\subsection{Parallel Grid Design}
\label{sec:pgrid}
\begin{figure}
	\centering
		\includegraphics[height=2cm]{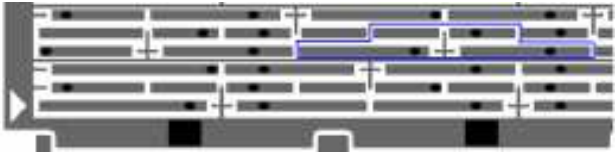}
	\caption{Geometry of the parallel bias-grid design. Highlighted is the symmetry cell.}
	\label{fig:pgrid}
\end{figure}

As shown in figure \ref{fig:pgrid}, the parallel grid implements a bias--grid line parallel
to the long direction of the sensor cells every third row. The intermediate rows are bound to 
the bias--grid by a punch--through dot in the middle of the pixel. 
The symmetry cell of this design is highlighted in figure \ref{fig:pgrid}.

In figure \ref{fig:pgrid}, one can also see the solution that was chosen for the outer edges
of a module. 
Not to have an increased capacitive coupling to the sensors outer rim, the sensor cell size
was set to 5/4 for the even and 3/4 for the odd rows of the first column, respectively.
This is only necessary at the outer edge of the sensor, not between the front-end chips.
In laboratory measurements, it was observed that these enlarged (reduced) cells shows a mean 
equivalent noise charge\footnote{The equivalent noise charge is defined as the standard deviation
of the error function that is observed by scanning the threshold of the discriminator.}
of 220 e$^-$ (long) and 180 e$^-$ (short), respectively. This is to be compared to the 
noise of 200e$^-$ for the uniform pixel size, and demonstrates that the noise is dominated by the 
sensors capacitance and not by the feed-throughs.

\subsection{Zig-Zag Grid Design}
\label{sec:zzgrid}
\begin{figure}
	\centering
		\includegraphics[height=2cm]{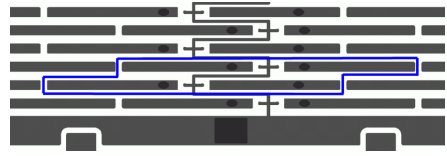}
	\caption{Geometry of the zig-zag bias-grid design. Highlighted is the symmetry cell.}
	\label{fig:zzgrid}
\end{figure}

A different approach to implement the bias-grid is shown in figure \ref{fig:zzgrid}. 
Here, the bias-grid follows the bricked structure and all bias dots are 
at the short edges of the sensor cells.

Both designs where implemented and tested. In the following sections a comparism 
of the two with respect to charge collection and spatial resolution will be given.
The principle ideas of the measurements enabled by data taken in test beam environment and 
about the set-up used can be found in \cite{NimAttilio}.

\section{Charge Collection}
\label{sec:chargecollection}
\begin{figure}[ht]
	\centering
		\includegraphics[clip=true,width=\columnwidth]{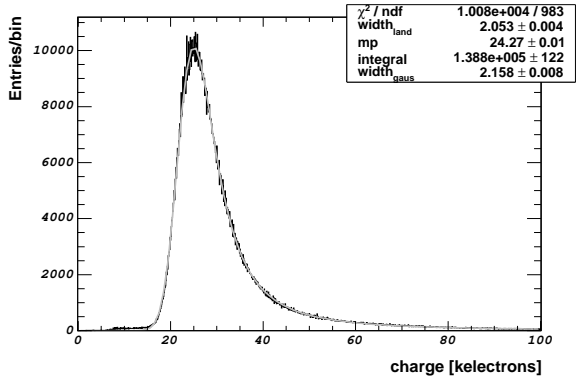}
	\caption{Charge of single hits. }
	\label{fig:allhitcharge}
\end{figure}

Although a binary readout with a discriminator is implemented in the ATLAS pixel electronics,
a so-called ``Time over Threshold'' mechanism provides information about the charge collected
by a sensor cell. In figure \ref{fig:allhitcharge}, the charge collected by single hits is given.
The fitted Landau distribution, folded with a Gaussian to cope for the noise sources, 
describes the data very well. The most probable charge is determined as $24,270\pm10\mathrm{e}^-$. 
The error does not account for the possible error in the calibration, 
which can well explain the discrepancy to the expected $22,400 \mathrm{e}^-$
in a $280\mu\mathrm{m}$ thick sensor. 
The tail to lower charges is not described by the function and is explained by charge losses to 
adjacent cells which did not reach the threshold.

\subsection{Charge Collection Maps} 
\begin{figure*}[tb]
	\centering
	\subfigure[Parallel grid]{
		\includegraphics[clip=true,width=\columnwidth]{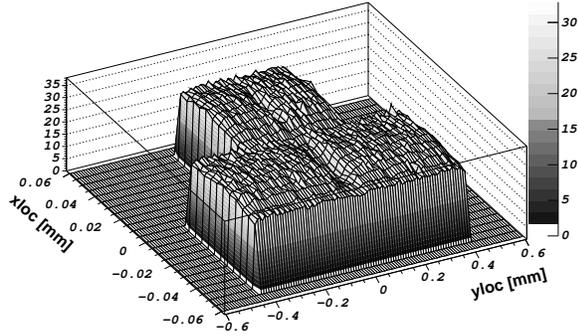}
		\label{fig:pg_chargemap}
	}
	\hfill
	\subfigure[Zig-zag grid]{
		\includegraphics[clip=true,width=\columnwidth]{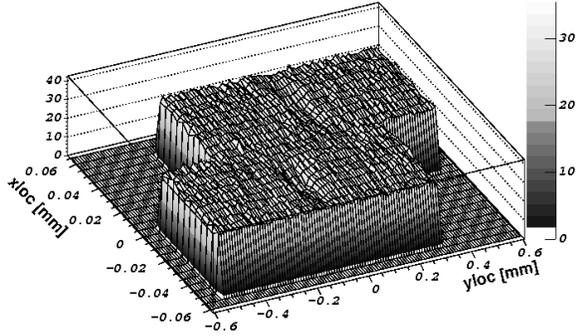}
		\label{fig:zz_chargemap}
	}
	\caption{Charge collection maps.}
	\label{fig:chargemaps}
\end{figure*}

Figure \ref{fig:chargemaps} shows the charge collection maps of the symmetry cells of a 
parallel grid and a zig-zag grid sensor. Here, all the charge seen by the cluster is summed up 
and plotted versus the hit prediction of the track defining telescope.
One can observe a slightly higher charge loss for the parallel grid. 
We will see later on that this is at least partly explained by the different threshold of the two
devices. 
It is to be pointed out that no variation of the charge collection could be observed
along the columns of the device. This means that there is no additional charge loss due to the 
increasing routing distances towards the middle axis of a module.

\subsection{Charge Collected}
The sum of the charges seen by the pixels in a cluster gives
a lower limit of the total charge liberated in the sensor. 
Due to the threshold applied by the discriminator in each pixel cell, 
some fraction of charge may be unreported in case of unequal charge sharing. 

 \begin{figure}
	\centering
	\subfigure[standard sensor]{
		\includegraphics[clip=true,width=\columnwidth]{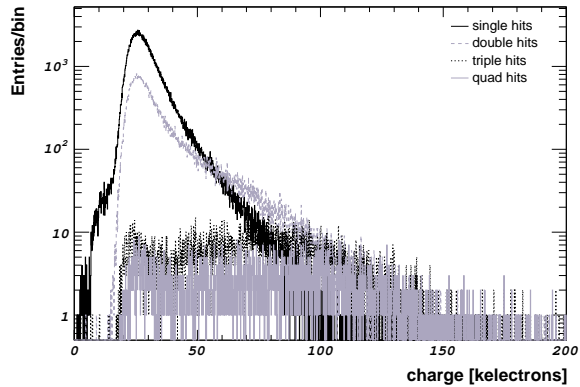}
		\label{fig:charge_std}
	}
	\subfigure[equal-sized-bricked sensor with parallel grid]{
		\includegraphics[clip=true,width=\columnwidth]{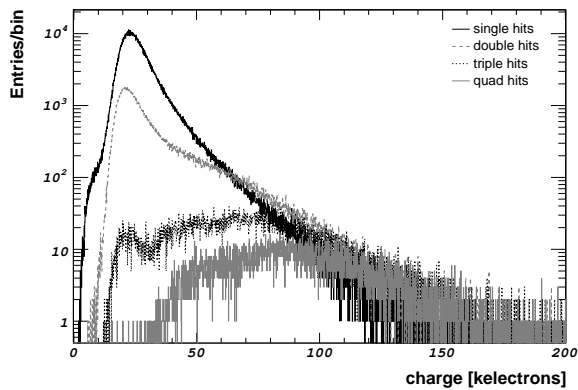}
		\label{fig:charge_pg}
	}
	\subfigure[equal-sized-bricked sensor with zig-zag grid]{
		\includegraphics[clip=true,width=\columnwidth]{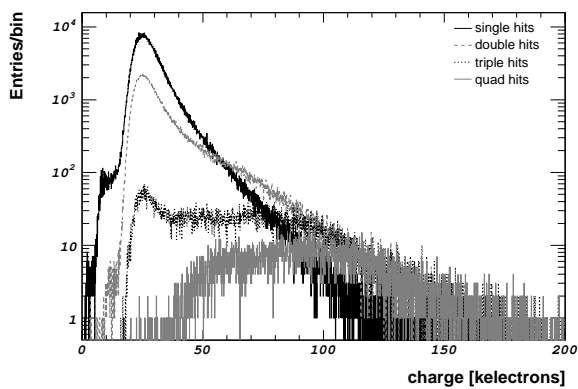}
		\label{fig:charge_zzg}
	}
	\caption{Charges of different cluster sizes for standard and 
		equal-sized-bricked sensors with zig-zag and parallel bias-grid. }
	\label{fig:charges}
\end{figure}

In figure \ref{fig:charges}, the summed charges of one to four pixel 
cluster sizes are shown for different sensor geometries.
For reference, the plots are shown for a standard sensor in figure
\ref{fig:charge_std}. This sensor has neither bricking nor equal sized pixels.
This can be seen in the fact that triple and quadruple 
clusters appear in the same shape.

Both have a Landau peak caused by minimum ionizing particles and a broad 
peak at higher charges, caused by higher energy depositions.
For the single hits, one can observe a tail towards lower values, 
which is caused by charge (below threshold) lost to neighbouring cells.

If we now compare this with figure \ref{fig:charge_pg}, which is the charge collected 
by an equal sized bricked sensor with parallel grid, one observes that the Landau peak 
for the quadruple hits has disappeared.
A bricked geometry does not allow a minimal ionizing particle to deposit energy
in four sensor cells. Instead, the Landau peak of the triple hit clusters increased.
Especially for limited operation conditions (like partly depleted sensors), 
this leads to an improved efficiency. 

One can also see an increase in the tail of low charges for single hits.
This is due to the fact the the threshold of this device was tuned to $5000 \mathrm{e}^-$.
This threshold is rather high, which causes some visible fraction of lost charge,
but the tracks are registered, because there is no charge sharing between four, 
but only between three cells.  

This can be confirmed by inspecting the figure \ref{fig:charge_zzg}. 
This shows the cluster charges observed in an equal sized bricked sensor 
with a zig-zag bias grid. 
The thresholds where tuned to $3000 \mathrm{e}^-$.
As expected, the tail of low charged single hits reduces, turning into 
double and mostly triple hits. The Landau-peak is now clearly visible for the triple hits.

We made some assumptions here regarding the constraints on charge distribution 
given by the sensors geometry. 
These can be confirmed by looking at the track positions for the different cluster sizes,
which will be done in the next section, together with the analysis of the spatial resolution.
\section{Spatial Resolution}
The spatial resolution may be determined by the residual of the centre of a cluster of pixels 
with respect to the prediction of the telescope. The telescope used has a track prediction uncertainty
of $5\mu\mathrm{m}$ within the pixel plane.
\subsection{Location Maps}
\begin{figure}[ht]
	\centering
		\includegraphics[clip=true,width=\columnwidth]{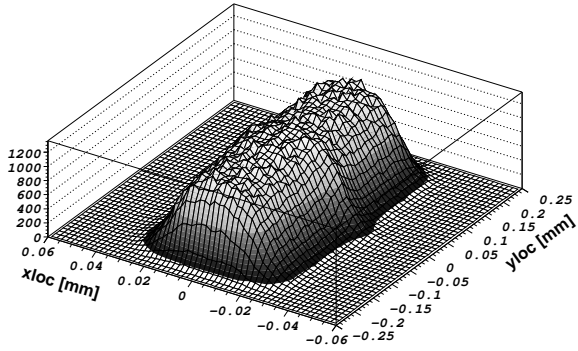}
	\caption{Location of single hits.}
	\label{fig:locmap_sgl}
\end{figure}

In figure \ref{fig:locmap_sgl}, the hit locations of tracks causing a single hit
are shown within the cell fired. One can observe the ``noses'' in the middle of the long edges,
where the three pixel area is situated.
%
\begin{figure}
	\centering
	\subfigure[Two hit cluster]{
		\includegraphics[clip=true,width=\columnwidth]{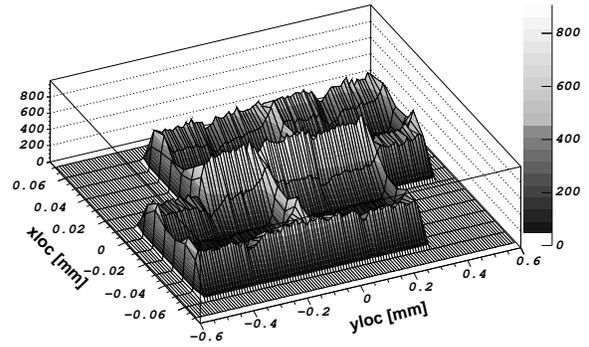}
		\label{fig:dbl_locmap}
	}
	\subfigure[Three hit cluster]{
		\includegraphics[clip=true,width=\columnwidth]{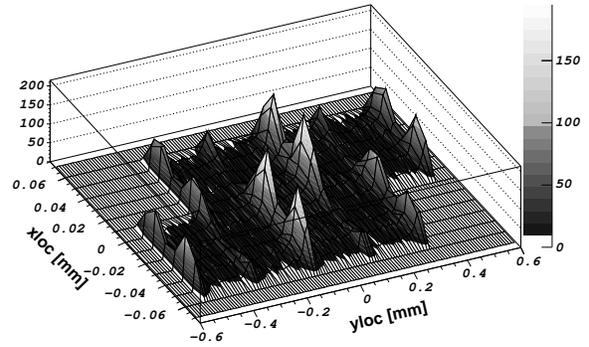}
		\label{fig:trp_locmap}
	}
	\subfigure[Four hit cluster]{
		\includegraphics[clip=true,width=.65\columnwidth]{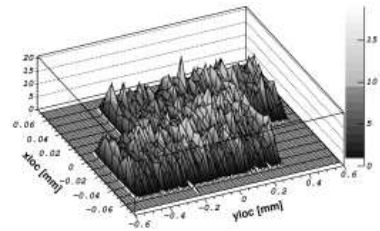}
		\label{fig:qd_locmap}
	}
	\caption{Location of tracks causing given cluster sizes, 
		shown here for the equal sized bricked sensor with zig-zag bias grid.}
	\label{fig:locmaps}
\end{figure}
The corresponding plots for two, three and four pixel cluster are given in figure \ref{fig:locmaps}.
For the two and three pixel cluster, the behaviour is as expected. 
One can clearly see the edges of the cells and the areas, where hits are causing three hit clusters. 

For the four pixel clusters, no clear accumulation can be observed. This is due to that fact that 
four pixel clusters need a large charge to be deposited and spread over an appropriate area (see above).
The same is true for part of the triple hit clusters (fig. \ref{fig:trp_locmap}), 
where one can observe sharp peaks from the minimum ionising particles in the regions of three cell edges.
In addition, there is a ``floor'' of non-located hits caused by large energy depositions.
\subsection{Residuals}
The Residuals are given by the differences between the geometrical centres of the clusters 
and the track positions, as determined by the telescope.
The TOT information (see section \ref{sec:chargecollection}) can be used to improve the 
information given by the pixels in case of multiple hits. 
For the the results shown in this section, only the geometrical information is used.

\begin{figure}
	\centering
	\subfigure[Standard Sensor]{
		\includegraphics[clip=true,width=\columnwidth]{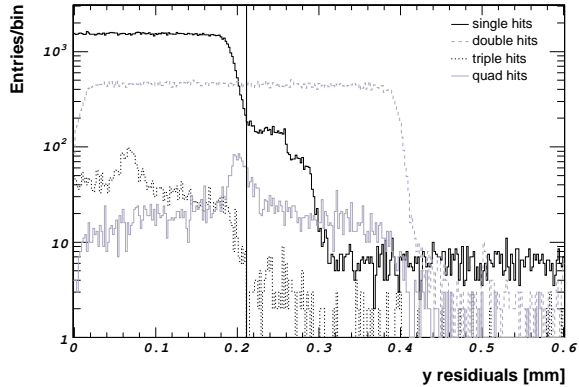}
		\label{fig:yres_std}
	}
	\subfigure[Equal sized bricked sensor with parallel bias grid]{
		\includegraphics[clip=true,width=\columnwidth]{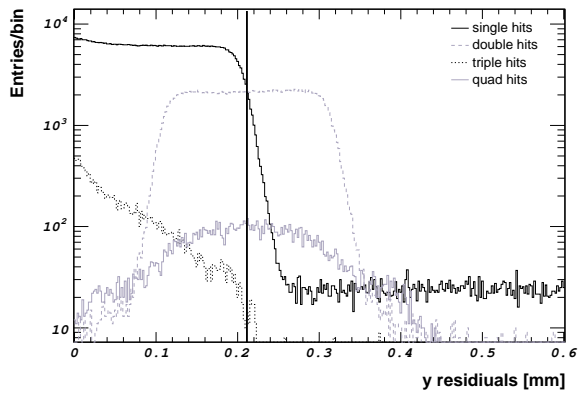}
		\label{fig:yres_pg}
	}
	\subfigure[Equal sized bricked sensor with zig-zag bias grid]{
		\includegraphics[clip=true,width=\columnwidth]{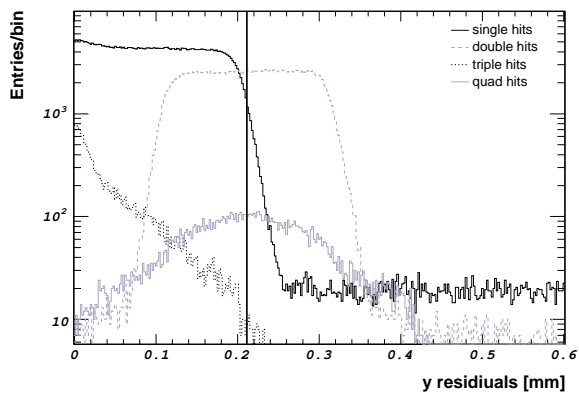}
		\label{fig:yres_zzg}
	}
	\caption{Residuals in the long direction of the sensor cells. Double and quadruple centres are shifted by half of a cell-length}
	\label{fig:yres}
\end{figure}

The analysis of the short direction is not discussed here. 
This is because there is no interesting structure visible for the equal sized sensor. 
(As it is for the standard sensor, featuring coupled cells to cover the middle axis of the module.)
This is different for the long direction of the cells, as shown in figure \ref{fig:yres}.

For better visualisation, there are shifts introduced in all plots shown.
The double and quadruple hit curves are shifted by half of the pixel length.
For double hits, only those of a single column are accepted. 
This suppresses the expected bump near zero (shown here at half pixel length), 
caused by two hit clusters of pixels facing their short edges.

For the standard sensor, the single hit curve shows a plateau of the usual cell length of $400 \mu\mathrm{m}$.
There is a step extending this to $600 \mu\mathrm{m}$, caused by the elongated cells to cover the inter-chip regions. 
Due to an artifact of this analysis, there are some of these hits suppressed, causing this step to be twofold.
The double hits don't show any structure. There is to notice the length of the plateau, which is the full cell length.
The triple hits show the broad distribution of high charge hits and the peak of the minimum ionizing particles.
This is shifted by a third of the pixel length, because these clusters consist of two pixels in one row and 
one in the neighbouring row. This shifts the geometrical centre.
Similarly, for the quadruple clusters, there is a broad background and a peak around the prediction.
This is because in this non-bricked design, there is a region where four cell edges are located.

The residuals for the equal sized bricked designs are much cleaner (figure \ref{fig:yres_pg} and \ref{fig:yres_zzg}).
There is no peak in the quadruple cluster location, 
and the peak for the triple hits is more pronounced, as seen in the
charges collected shown in figure \ref{fig:charges}.
There is a small peak visible for the single hit, 
caused by the ``noses'' shown in figure \ref{fig:locmap_sgl}.
For the double hits, the improved resolution of half a pixel length is clearly visible.
The width of this plateau is $208 \mu\mathrm{m}$, 
which is to be compared to the cell length of $422.22\mu\mathrm{m}$.

\section{Summary}
The performance of equal sized bricked sensors making use of the possibilities
offered by the use of the MCM-D technique was demonstrated in a test beam environment.
The charge collection is not degraded, neither by the sensor geometry nor by the feed-through structures.
This geometry does not allow for low charges induced in the sensor to be shared by more than three
pixel cells, which yields an improved efficiency.
The cleaner residual plots eases the tracking algorithms, needing less distinction of cases, 
and the spatial resolution for double hits is improved by a factor of two for the long direction
of the cells. The results presented here are part of a thesis \cite{ChGDis}, 
where more detailed information can be found. 

\section{Acknowledgements}
This work was supported by the ATLAS Pixel project and 
has been granted by the German Federal Government BMBF 
under project numbers 05\,HA8PXA\,2 and 05\,HA1PX1\,9.
 
The authors especially want to thank all participants of the test beam campaign.
We also want to thank the members of the Fraunhofer Institut IZM for their engagement
in this corporate project.




\begin{thebibliography}{00}
\bibitem{mcmdpixel}
    C.~Grah et al.,
    \emph{Pixel Detector Modules using MCM--D technology}
    Nucl. Instr. and Meth. A 465 (2001) 211--218
\bibitem{becks}
  K.H.~Becks et al.,
    \emph{A MCM--bD type module for the ATLAS Pixel Detector}
    Proc. of IEEE Nuclear Science Symposium, Toronto, Canada 1998
\bibitem{NSS04}
 K.H.~Becks et al.,
    \emph{Building Pixel Detector Modules in Multi Chip Module Deposited
  Technology}
    Proc. of the IEEE Nuclear Science Symposium, Rome, Italy 2004
	\note also submitted to Transaction on Nuclear Science
\bibitem{Siena02}
  T.~Flick, K.H.~Becks, P.~Gerlach, C.~Grah, P.~M\"{a}ttig and T.~Rohe,
  \emph{Studies on MCM-D Pixel-Detector-Modules}
  Published in Nucl. Phys. B (proc. suppl.) volume 125,
  8th topological seminar on innovative particle and radiation detectors,
  Siena 2002
\bibitem{Pixel02}
  K.H.~Becks, T.~Flick, P.~Gerlach, C.~Grah, P.M\"{a}ttig
  \emph{Studies on MCM-D interconnections}
  Contribution to Pixel2002 International Workshop on Semiconductor Pixel
  Detectors for Particles and X-Rays, Carmel 2002
\bibitem{NimATLAS}
 M.S.~Alam, et al.
 \emph{The ATLAS silicon pixel sensors}
 Nucl. Instr. and Meth. A 456 (2001) 217--232
\bibitem{NimAttilio}
  A.~Andreazza
  \emph{Performance of ATLAS pixel detector prototype modules}
  Nucl. Instr. and Meth. A 513 (2003) 103--106
\bibitem{ChGDis}
  C.~Grah
  \emph{Development of the MCM-D Technique for Pixel Detector Modules}
  Bergische Universi{\"a}t Wuppertal; WUB-DIS 2005-05;\\ 
  urn:nbn:hbz:468-20050296
  
  




\end{thebibliography}
\end{document}